**Imaging ballistic carrier trajectories in graphene using scanning gate microscopy**

Sei Morikawa[1], Ziwei Dou[2], Shu-Wei Wang[2], Charles G. Smith[2], Kenji Watanabe[3], Takashi Taniguchi[3], Satoru Masubuchi[1], Tomoki Machida[1,4, a] , and Malcolm R. Connolly[2,b]

[1]*Institute of Industrial Science, University of Tokyo, 4-6-1 Komaba, Meguro, Tokyo 153-8505, Japan*
[2]*Cavendish Laboratory, Department of Physics, University of Cambridge, Cambridge CB3 0HE, UK*
[3]*National Institute for Materials Science, 1-1 Namiki, Tsukuba 305-0044, Japan*
[4]*Institute for Nano Quantum Information Electronics, University of Tokyo, 4-6-1 Komaba, Meguro, Tokyo 153-8505, Japan*

**Abstract:**

We use scanning gate microscopy to map out the trajectories of ballistic carriers in high-mobility graphene encapsulated by hexagonal boron nitride and subject to a weak magnetic field. We employ a magnetic focusing geometry to image carriers that emerge ballistically from an injector, follow a cyclotron path due to the Lorentz force from an applied magnetic field, and land on an adjacent collector probe. The local electric field generated by the scanning tip in the vicinity of the carriers deflects their trajectories, modifying the proportion of carriers focused into the collector. By measuring the voltage at the collector while scanning the tip, we are able to obtain images with arcs that are consistent with the expected cyclotron motion. We also demonstrate that the tip can be used to redirect misaligned carriers back to the collector.

[a] E-mail: tmachida@iis.u-tokyo.ac.jp
[b] E-mail: mrc61@cam.ac.uk

Recent advances in techniques for transferring and stacking single atomic layers have enabled the fabrication of van der Waals heterostructures of graphene and hexagonal boron nitride (h-BN) [1–4]. Graphene encapsulated by h-BN exhibits particularly high carrier mobility of $\mu > 100,000$ cm$^2$/Vs [2] because, unlike previous generations of devices with graphene on SiO$_2$, extrinsic carrier scattering from charged impurities, surface roughness, and optical phonons of SiO$_2$ is largely eliminated [3]. Carriers consequently travel ballistically over a mesoscopic distance, of the order of several microns in graphene/h-BN, allowing the observation of transport phenomena such as negative bend resistance [5], magnetic electron focusing [6], and an anomalous magnetoresistance peak due to boundary scattering [7]. While there is a qualitative similarity with these effects in conventional two-dimensional electron systems (2DESs) based on semiconductor Al$_x$Ga$_{1-x}$As/GaAs heterostuructures [8–10], hallmarks of the linear Dirac spectrum of excitations in graphene manifest through room-temperature ballistic carrier transport [5], Klein tunneling [11], the formation of snake states [12,13], and an anomalous commensurability ratio at the magnetoresistance peak due to boundary scattering [7].

Scanning gate microscopy (SGM) is now a well-established technique for visualizing local variations in the electronic properties of 2DESs [14–17]. SGM involves monitoring the conductance of the 2DES while scanning a sharp metallic tip over its surface. A bias voltage applied between the tip and 2DES modifies the carrier density under the tip, locally shifting the Hartree potential and Fermi level. In disordered low-mobility systems, the conductance predominantly changes due to the carrier density-dependent diffusion constant [18], while the behavior of ballistic high-mobility devices is more accurately described in terms of diffraction of electron rays by tip-induced changes in the Fermi wavelength, a fact that has been employed in GaAs 2DESs to reveal ballistic carrier trajectories during magnetic electron focusing [19].

In this letter, we demonstrate how SGM can be used to reveal ballistic carrier trajectories in high-mobility graphene encapsulated by h-BN. We concentrate on capturing SGM images of excitations that are injected from a contact in a magnetic field and travel directly from injector to collector along a curved cyclotron orbit. This phenomenon, known as focusing due to the fact that electron density becomes singular on the caustic of carrier trajectories [6], is detected by measuring the non-local voltage $V_{NL}$. This voltage is only generated when the cyclotron diameter of the charge carriers fits an integer number of times into the distance between the injector and collector probes [6,8]. In SGM the local electric field generated by the tip perturbs the cyclotron motion of the ballistic carriers [20], causing them to miss the collector and lowering the spatial charge build up responsible for $V_{NL}$. Since the greatest disturbance from the tip occurs when it is directly over the beam of carriers, SGM images are characterized by arc-shaped areas where the reduction in $V_{NL}$ is most pronounced.

For this study we fabricate a Hall-bar device on graphene encapsulated by h-BN crystals using a mechanical exfoliation and "pick-up" transfer technique of atomic layers [2]. The assembly of graphene and h-BN layers in the present work utilizes van der Waals forces to pick up the exfoliated atomic layers from the surface of the Si substrate with a 290-nm-thick thermally oxidized layer of $SiO_2$ [2]. Compared to the conventional technique for transfering atomic layers using a polymer film [3,4], the polymer-free assembly technique of atomic layers has the advantage that the interface between graphene and h-BN does not come into contact with any polymer film throughout the entire process. This reduces the contamination of the channel and associated degradation of the carrier mobility. The sample consists of ~ 50 nm of h-BN on the bottom of a mono-layer graphene and ~ 50 nm of h-BN as a cap layer. The sample was masked with a Hall-bar-shaped hydrogen silsesquioxane (HSQ) resist and plasma-etched to expose the edge of the graphene [2]. Finally, electron-beam lithography, evaporation of Cr/Pd/Au metal, and a subsequent lift-off process were used to form one-dimensional contacts to the graphene edges [2]. The four-terminal longitudinal resistance measured as a function of the back-gate-bias voltage $V_{bg}$ at the measurement temperature $T$ = 4.2 K is shown in Fig. 2(a), revealing a small residual doping of $\Delta n$ ~ $10^{11}$ cm$^{-2}$ and a high charge-carrier mobility of 60,000–80,000 cm$^2$/Vs. The mean free path is estimated as ~ 1 μm at $n = 10^{12}$ cm$^{-2}$, which is one order of magnitude larger than that for graphene on SiO$_2$ and is comparable to the size of the graphene channel.

Our scanning gate microscope is mounted inside a cryostat with a base temperature of ~ 4.2 K. The oscillation of the cantilever was measured using standard interferometric detection with a fiber-based infra-red laser. We used a conductive diamond-coated cantilever (NanoWorld POINTPROBE-CDT-FMR-10) with a nominal tip radius of 100–200 nm. In order to avoid any cross-contamination between the tip and the flake during SGM, we mostly operate in lift-mode with the static tip at a height of ~ 10 nm from the top h-BN surface. Based on electrostatic simulations, the FWHM of electric field broadening from the SGM tip is thus expected to be ~ 200 nm (details are discussed in Sec. A in the supplementary information [21].). Broadening at this scale is small enough to image the local ballistic carrier trajectories during focusing (See Fig. S1 in the supplementary information [21]). In order to associate the SGM response with specific locations of the device, we first measure the resonant frequency shift $\Delta f$ of the SGM tip while applying a small $V_{bg}$. Since $\Delta f$ depends on the screening of the electric field from the back gate, maps of $\Delta f$ reveal the positions of the graphene and the contacts. Comparing the map of $\Delta f$ with an optical microscopy image of the device then enables us to identify the area to be scanned without degrading the tip.

First, we use conventional carrier transport measurements to confirm the ballisticity of excitations via the magnetic focusing effect in a weak magnetic field. The measurement configuration is shown in Fig. 1. With the tip parked far from the device, we inject carriers using a current $I = 100$ nA from contact i to contact g and measured $V_{NL}$ between contact c and contact f using conventional lock-in techniques. Contact i and contact c work as an injector and a collector of carriers, respectively. Note that $V_{NL}$ should be zero if the carriers travel diffusively because the voltage probes are located far from the path of charge carriers scattered by random impurities.

Figure 2(b) shows $V_{NL}$ as a function of the magnetic field $B$ at $V_{bg} = 10$, 15, and 25 V at $T = 4.2$ K. We observe two distinct peaks in $V_{NL}$ whose position in magnetic field, tracked by red and blue arrows, increases with $V_{bg}$. This behaviour is consistent with the earlier work by T. Taychatanapat *et al.*, where the magnetic focusing of Dirac fermions in graphene was reported and can be explained as follows [6]. In a weak magnetic field, carriers emerging from the injector contact travel ballistically along a trajectory that is curved by the Lorentz force. The trajectory corresponds to a classical path of carriers performing cyclotron motion with radius, $r_c = \hbar\sqrt{n\pi}/eB$, where $n$, $\hbar$, and $e$ denote the carrier density, Planck's constant, and elementary charge, respectively. At magnetic fields corresponding to the peaks indicated by the red solid arrows, the diameter of cyclotron motion $2r_c$ is equal to the distance between the injector and collector $W_{IC}$, i.e., $2r_c = W_{IC} = 750$ nm, and the ballistic carriers reach the collector contact directly, resulting in a positive value of $V_{NL}$ [Fig. 2(c)]. At the magnetic field where the peak indicated by the blue arrow is observed, $2r_c = W_{IC}/2$, carriers first bounce off the edge before being detected at the collector contact, as shown in Fig. 2(d). $V_{NL}$ should be positive for this case as well. To confirm that the peaks in $V_{NL}$ are observed when either the condition $2r_c = W_{IC}$ or $2r_c = W_{IC}/2$ is satisfied, we compared the values of the peak magnetic field obtained experimentally with those expected for the cases when $2r_c = W_{IC}$ and $2r_c = W_{IC}/2$, as shown in Fig. 2(e). The red and blue circles indicate the first (lower $B$) and the second (higher $B$) peak magnetic field obtained experimentally. The peak position is determined by fitting a Gaussian profile. The red and blue lines correspond to the estimated magnetic fields for Figs. 2(c) and 2(d), respectively. This can be expressed as

$$B = \left( \frac{2\hbar\sqrt{\pi n}}{eW_{IC}} \right) p \ ,$$

where $p = 1, 2$ (red, blue) [6]. $n$ is derived from $n = C_g(V_{bg} - V_{Dirac})/e$, where $C_g$ is the capacitance between channel and back gate determined from quantum Hall measurements, and $V_{Dirac}$ is the back-gate-bias voltage corresponding to the charge neutrality point. The uncertainity in the theoretical estimations are indicated by the width of the lines. The excellent agreement between the two suggests that the observed peaks are indeed signatures of magnetic focusing and hence confirms that carriers travel ballistically in our device. Note that the specularity of the sample edge can be estimated as $s \sim 0.45$–$0.65$ from the amplitude ratio of the first and the second peaks [6,22].

By combining the picture of magnetic focusing described above with SGM measurements we are able to reveal ballistic trajectories of Dirac fermions. Figure 3(a) shows the $V_{NL}$ obtained when the SGM tip is scanned with an applied tip bias of $V_{tip} = -20$ V at $B = 0.22$ T and $V_{bg} = 10$ V. These $B$ and $V_{bg}$ values correspond to the first focusing peak shown by a red solid arrow at $V_{bg} = 10$ V in Fig. 2(b), where bulk transport measurements suggest carriers are focused into the collector. The expected position of the sample edge indicated by a red line is estimated from the $\Delta f$ measurements as discussed above. A suppressed $V_{NL}$ signal is observed in an arc-shaped (black) region spanning the injector and collector voltage probes (The wider scan frame image is shown in Fig. S3 in the supplementary informaton [21]). A line section of $V_{NL}$ when the tip is scanned along the white dotted line in Fig. 3(a) is shown in Fig. 3(e). The suppressed $V_{NL}$ signal can be understood as follows. When the tip is on or near the ballistic carrier cyclotron trajectory, the local electric field generated by the tip deflects the carriers [20] from the circular paths connecting the injector and collector, as schematically depicted in Fig. 3(d), where the solid and dotted arrows depict carrier trajectories with and without the tip, respectively. The proportion of carriers reaching the collector contact consequently reduces, causing a reduction in $V_{NL}$. Intuitively, we expect the precise magnitude of $V_{NL}$ to depend on the extent of the deflection and the density of the carriers disturbed by the tip. The arc-shaped suppressed $V_{NL}$ region in Fig. 3(a) is thus a direct impression of the original ballistic trajectories taken by carriers travelling between the contacts. We note that broadening of arc-like features can be interpreted as the effect of the finite width of the injector and the angle distribution of the injected carriers. (For further details see Sec. C in the supplementary information [21].) Clear confirmation of this picture is shown in Fig. 3(b), when we expect ballistic carriers to be first reflected by the edge before being picked up at the collector probe, as depicted in the accompanying schematic in Fig. 2(d). Fig. 3(b) shows spatial mapping of $V_{NL}$ obtained when the SGM tip is scanned at $B = 0.46$ T and $V_{bg} = 10$ V, corresponding to the second focusing peak shown by a blue arrow at $V_{bg} = 10$ V in Fig. 2(b). As anticipated, this time we find two arc-shaped suppressed $V_{NL}$ regions between the injector probe and collector probe. The line cut of this image along the dotted line [Fig. 3(g)] shows three dips in $V_{NL}$ (indicated by black arrows), confirming the presence of two arc-shaped suppressed $V_{NL}$ regions. Using the same argument as above, these structures are entirely consistent with the original ballistic carrier trajectory [Fig. 3 (f)] and confirms that electrons bounce once from the edge in this regime [Fig. 2(d)]. Note that $V_{NL}$ is actually enhanced when the tip is near the edge of the graphene in Figs. 3(a) and 3(b), possibly due to the way caustics are distorted by the electrostatics at the edge and tip-induced refocusing of carriers into the collector [23].

Next we set $B = 0.26$ T, a field slightly greater than the focusing field [red dotted arrow at $V_{bg} = 10$ V in Fig. 2(b)]. $V_{NL}$ is less than the peak value, meaning that the focal point of trajectories falls on the injector side of the collector contact [Fig. 3(h)]. SGM images with $V_{tip} = 10$ V exhibit arc-shaped regions but with enhanced $V_{NL}$ (white) bridging between the injector and collector probe, as shown in Fig. 3(c). The line cut of this image along the white dotted line [Fig. 3(i)] confirms the enhanced value of $V_{NL}$ around the focusing region. When the tip is within the arc-shaped region, it deflects carriers just enough to deflect them back into the contact despite the Lorentz force on its own being insufficient for focusing. The enhanced $V_{NL}$ region in Fig. 3(c) shows the positions where the tip can realign the ballistic carriers to the collector probe, as shown in Fig. 3(h). This suggests the tip is able to locally reconfigure the focal point of ballistic trajectories. Since the carrier trajectories that initially fall on the injector side of the collector are shifted toward the collector probe by applying a positive $V_{tip}$, the tip-induced realignment of ballistic carrier trajectories is explained qualitatively by the local increase of the cyclotron radius $r_c \propto \sqrt{n}/B$ induced by the increase in the local carrier density under the tip. However, the underlying mechanism responsible for the deflection itself requires further investigation. To shed some light on the deflection mechanism we parked the static tip within the focusing region and performed magnetic field spectroscopy. As discussed in Sec. D in the supplementary material [21], the behavior of the peak position at different tip bias is consistent with the description above. More detailed discussion about the characterization of tip influence is also shown in Sec. E in the supplementary material [21].

In summary, we have reported the imaging of ballistic carrier trajectories in graphene using SGM and high-mobility graphene encapsulated by h-BN crystals. We imaged ballistic carrier trajectories in graphene by deflecting the trajectories using a scanning tip. Finally, we showed how to use the tip to intentionally move the focal point of trajectories to a specific location in a device. Our results show the potential for using SGM to understand how Dirac fermions behave in ballistic quantum devices. For example, further study of the reflection of carriers at the edges should allow a better understanding of edge scattering. Moreover, our results also demonstrate that scanning probe tips are suitable as mobile lenses for manipulating ballistic carriers. Electron optics [24,25] in graphene continues to inspire device concepts not possible with conventional 2DESs, such as Veselago lensing [26], wave guiding [27], and cloaking [28], and our results offer a direct insight into how tips might be used to prototype devices and perform experiments that require manipulation of electron flow on a local scale [29].

After completion of this study, we became aware of a related work with similar results of imaging the magnetic focusing effect by S. Bhandari *et al* [23].

**Acknowledgments**

S. Morikawa and Z. Dou contributed equally to this work. The authors acknowledge M. Arai and R. Moriya for technical support, simulation of electric field broadening using finite element method, and discussions. This work was partly supported by the EPSRC; a Grant-in-Aid for Scientific Research on Innovative Areas "Science of Atomic Layers" from the Ministry of Education, Culture, Sports, Science and Technology (MEXT); the Project for Developing Innovation Systems of MEXT; the Grants-in-Aid for Scientific Research from the Japan Society for the Promotion of Science (JSPS); and the CREST, Japan Science and Technology Agency. S. Morikawa acknowledges the JSPS Research Fellowship for Young Scientists and the Advanced Leading Graduate Course for Photon Science (ALPS), University of Tokyo. Additional data related to this publication is available at the University of Cambridge data repository (https://www.repository.cam.ac.uk/handle/1810/252877).

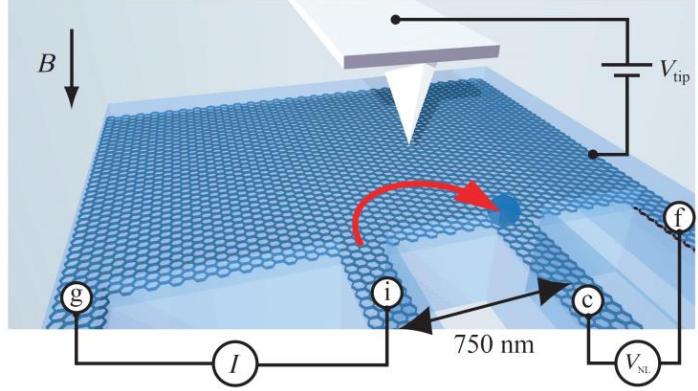

Fig. 1.

(Color online) Schematic of our setup used to perform scanning gate microscopy (SGM). A local electric field is generated near the tip by applying $V_{tip}$ to a conductive diamond-coated cantilever. The tip is scanned at a height of ~ 10 nm from the surface of h-BN/graphene/h-BN. A magnetic field $B$ is applied perpendicularly to the graphene. A current $I$ = 100 nA is injected from contact i to contact g, and the non-local voltage $V_{NL}$ is measured between contact c and contact f. The distance between the injector and collector probes is $W_{IC}$ = 750 nm.

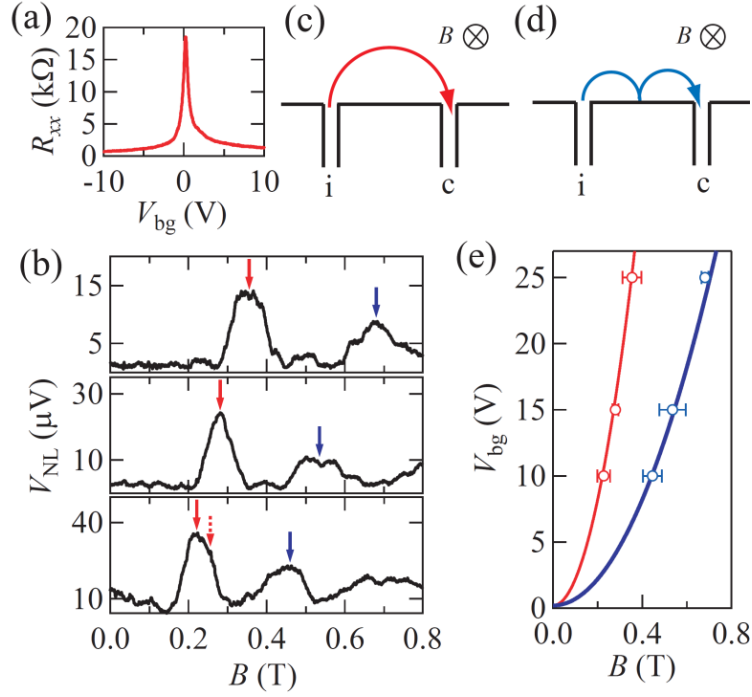

Fig. 2

(Color online) (a) Four-terminal longitudinal resistance $R_{xx}$ as a function of the back-gate-bias voltage $V_{bg}$ at $T = 4.2$ K. (b) Non-local voltage $V_{NL}$ as a function of the magnetic field $B$ for $V_{bg} = 25$, 15, and 10 V (top to bottom). Schematics of the magnetic focusing carrier trajectories when (c) $2r_c = W_{IC}$ and (d) $2r_c = W_{IC}/2$, where $r_c$ and $W_{IC}$ are the cyclotron radius and the distance between contact i and contact c. (e) Red and blue circles show the values of $V_{bg}$ v.s. $B$ at the first (lower $B$) and the second peak (higher $B$) of $V_{NL}$. Red and blue lines show $V_{bg}$ v.s. $B$ satisfying $2r_c = W_{IC}$ and $2r_c = W_{IC}/2$, respectively.

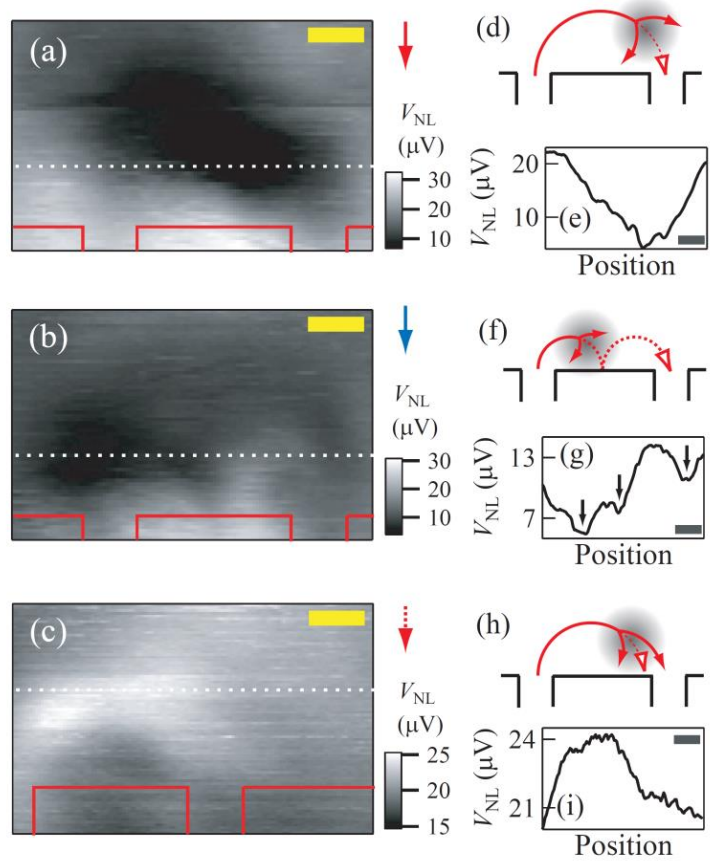

Fig. 3

(Color online) Scanning gate images showing the spatial variation of $V_{NL}$ at $V_{bg} = 10$ V for (a) $B = 0.22$ T and $V_{tip} = -20$ V, (b) $B = 0.46$ T and $V_{tip} = -20$ V, and (c) $B = 0.26$ T and $V_{tip} = 10$ V, corresponding to the red solid, blue solid, and red dotted arrow in Fig. 2(b), respectively. The red lines indicate the sample edges. The scale bar in each figure is 200 nm. Schematic of the carrier trajectory corresponding to (a), (b), and (c) are illustrated in (d), (f), and (h), respectively. The solid and dotted arrows depict carrier trajectories with and without the tip. The grey circles indicate the regions over which the carrier density is strongly modulated by the tip. Spatial distribution of $V_{NL}$ along the white dotted line in (a), (b), and (c) are shown in (e), (g), and (i), respectively. The arrows in (g) mark the dip in $V_{NL}$.